\documentstyle[11pt,newpasp,epsfig,twoside]{article}
\def\edcomment#1{\iffalse\marginpar{\raggedright\sl#1\/}\else\relax\fi}
\reversemarginpar

\newcommand{\cdo}{$^{12}C(\alpha ,\gamma)^{16}O\;$}

\begin{document}
\title{The star formation history of resolved galaxies: Composite H-R diagram.}
 \author{Carme Gallart}
\affil{Andes Fellow. Astronomy Dept. Yale University. P.O. Box 208101. New Haven, 
CT 06520-8101, USA {\rm and} Departamento de Astronom\'\i a, Universidad de Chile. Casilla 36-D. Las Condes. Santiago. Chile}

\begin{abstract}
I discuss the information provided by color-magnitude diagrams of different 
absolute magnitude limits in the study of the star formation histories of 
nearby galaxies. I review the recent progress made in the field, with 
emphasis on the results for Local Group dwarf galaxies.
\end{abstract}

\section{Introduction}
We can study the evolution of the nearest galaxies in exquisite 
detail from the information contained in the H-R diagram -or Color-Magnitude 
Diagram, CMD-  of their resolved stars. From it, it is possible to 
retrieve the star formation and chemical enrichment histories (SFH 
and Z(t)), with a level of detail that mainly depends on the magnitude limit
of the available CMD. For CMDs 
reaching the oldest main-sequence (MS) turnoffs, it is possible to break 
the age-metallicity degeneracy that affects most methods of retrieving
SFHs. Even if we can pursue this kind of study only in a very limited 
volume of the Universe, 
it is a worthwhile task, key to understand galaxy formation and
evolution in general, because: 

$\bullet$ In our neighborhood, we have representatives of almost any kind of 
galaxies in the Universe: in the Local Group there are three spirals, and a 
large number of dwarf -both dIr and dSph- 
galaxies; the nearest gE, NGC5128 is at $\simeq 4.0$  Mpc, and 
similarly, the nearest BCD and LSB galaxies are in the M81 group, also at 
$\simeq 4.0$ Mpc. In addition, by studying the details of the SFHs 
in the nearest galaxies, we may be able to identify the evolved 
counterparts of distant objects that are experiencing characteristic 
star forming episodes.

$\bullet$ This kind of study provides detailed information on 
the first events of star formation in the nearest galaxies. Therefore,
we can address questions like ``have the oldest stars in galaxies all the same age?'' or, ``are there differences, in the epoch or strength of the early
star formation, depending on environment''. 

$\bullet$ We can test the hypothesis of galaxies being assembled by merging 
through the almost direct observation of the merging history of the Milky 
Way.

$\bullet$ It offers an alternative to the cosmologic look-back-time 
approach to study galaxy evolution, by tracing the complete evolution of 
particular objects.

Furthermore, it is indeed a feasible task. From the observational point
of view, we have the means of gathering the necessary observations. For the
nearest objects, the dSph satellites of the Milky Way and the Magellanic 
Clouds, we have excellent
wide-field CCDs on ground-based telescopes. Observations of this kind have
already revealed in amazing detail the varied SFHs of these galaxies, and 
we will be able to reach the same level of detail for the remaining Local 
Group galaxies using the Advanced Camera for Surveys (ACS) on the 
HST. Further improvements in optical capabilities on space (e.g. NGST) or
the ground (e.g. adaptive optic systems) will be important to pursue the 
same kind of studies, with similar precision to larger distances. From the theoretical point of view, and
hand in hand with increasing computational power, we are seeing important 
improvement on stellar evolution models (through e.g. more realistic 
treatments of convection or rotation, or improved opacities), and on the 
accurate modeling of synthetic CMDs with realistic simulations of 
observational errors. 

There is three main levels of information in the CMD depending on the 
magnitude limit of the available data: the one we obtain when we reach 
the oldest MS turnoffs, the one provided by reaching the horizontal 
branch (HB) level, 
and when we only observe a couple of magnitudes below the tip of the 
red-giant branch (RGB); see Figure 1. I will discuss them separately below.

\section{A quantitative method of deriving SFHs: synthetic CMDs}

The comparison of the observed CMDs with synthetic CMDs computed using
stellar evolution models is a fundamental tool that 
is becoming widely used to {\bf quantitatively} retrieve the SFH from 
the CMD. With this technique, a number of synthetic -or model- CMDs 
are computed assuming 
possible scenarios for the SFH, which can be reasonably decomposed in 
a number of simpler functions: the instantaneous star formation rate, 
SFR($t$), the chemical enrichment law, $Z(t)$, the initial mass 
function, IMF, and a function $\beta(f,q)$ controlling the fraction 
$f$ and mass ratio distribution $q$ of binary stars (Aparicio 1998; 
Gallart et al. 1999b). This basic approach has been followed, with 
variation in the details, 
by a number of groups (Bertelli et al. 1992; Gallart et al. 1996a; 
Tolstoy \& Saha 1996; Dolphin 1997; Hern\'andez, Valls-Gabaud \& Gilmore 
1998; Hurley-Keller, Mateo \& Nemec 1998; 
Ng 1998), and it has proven particularly successful, 
and leading to basically unique results when applied to data reaching  
the oldest MS turnoffs (Gallart et al. 1999b).
In the following, I will comment on the application of this method to CMDs 
of different magnitude limits.

\section{The best case: when the CMD reaches the oldest MS turnoffs}

During the last decade, our conception of the evolution of the dSph 
galaxies satellites of the Milky Way, has 
changed dramatically from the idea that they were predominantly old 
systems, to our current 
knowledge of their varied SFHs. Indeed, we find almost 
every imaginable evolutionary history in this sample of galaxies, from 
the extreme case of Leo I (Gallart et al. 1999a,b), 
which has formed over 80\% of its stars from 6 to 1 Gyr ago, to 
intermediate cases like 
Carina (Smecker-Hane et al. 1996; Hurley-Keller et al. 1998) 
and Fornax (Stetson, Hesser \& Smecker-Hane
1997; Buonanno et al. 1999), with prominent intermediate-age populations, to 
predominantly old ($\simeq$ 10 Gyr old) systems like Sculptor 
(Hurley-Keller, Mateo \& Grebel 1999), 
Draco (Aparicio, Carrera \& Mart\'\i nez-Delgado 2000), Ursa Minor 
(Mart\'\i nez-Delgado \& Aparicio 1998) or Leo II (Mighell \& Rich 1996). 
Studies of the LMC SFH have also benefited of CMDs of this quality (e.g.
Holtzman et al. 1999). This clear and 
extremely interesting picture has only been obtained when CMDs 
{\bf reaching the oldest MS 
turnoffs}, and covering a substantial fraction of the galaxy have 
been available.

Even though CMDs of this kind offer {\bf qualitative} {\it first glance} SFHs 
(e.g. in the case of Carina,  the most emblematic example, one can 
{\it see}, just by comparing the CMD with isochrones, 
that there has been three major events of star formation), 
a {\bf quantitative} determination of the SFH requires a detailed 
comparison of the distribution of stars in the CMD with that predicted 
by model CMDs.  With CMDs reaching the old MS turnoffs, there is 
the advantage that the stellar evolution involved 
(that of MS and subgiant-branch stars) is relatively well 
known, and there is less intrinsic degeneracy age-metallicity. Indeed,   
we have shown that, with this method, it is possible to break the classical
age-metallicity degeneracy in stellar populations (Gallart et al. 1999b). 
This is so because 
detailed information on the age distribution of the stars is gained from
their distribution along the MS and subgiant branches. Once 
this is determined, the possibilities for metallicity distributions 
are relatively unique in order to fit the detailed distribution of stars 
in the CMD. Admittedly, the position of the stars in the MS and subgiant branches also depends on metallicity, but, in 
practice, only narrow combinations of $Z(t)$, SFR($t$) and
$\beta(f,q)$ can reproduce the distribution of stars in the CMD. 

\begin{figure}
\begin{center}
\mbox{\epsfig{file=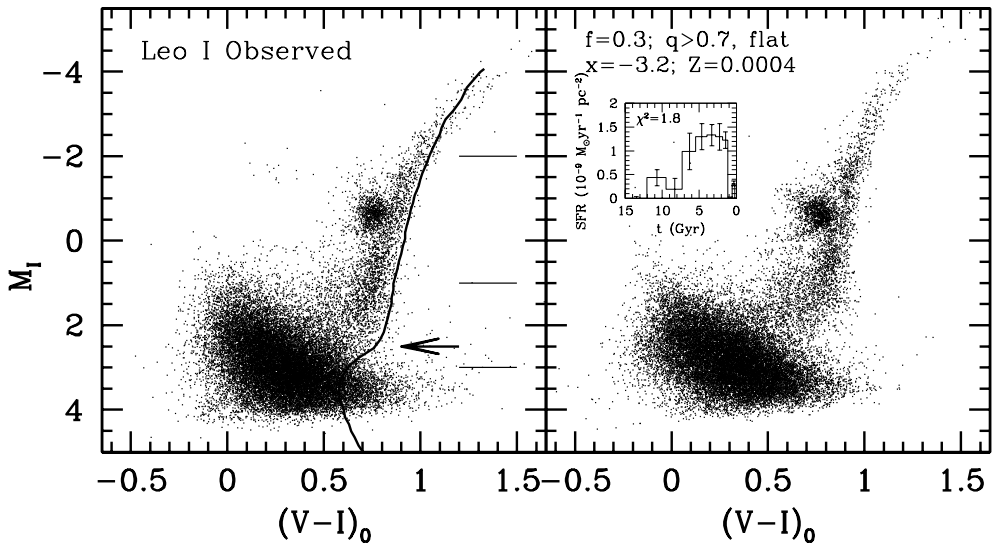,width=14cm,bbllx=0.5truecm,bblly=10.0truecm,bburx=21.0truecm,bbury=19.0truecm}}
\end{center}
\caption[]{Left panel: HST CMD of Leo~I. The three magnitude limits discussed 
in the text are shown by horizontal lines. A 15 Gyr old isochrone of Z=0.0004
from Bertelli et al. (1994) has been superimposed to the data, and the 
position of the old 
main-sequence turnoff is pointed by an arrow. Right panel: Summary of 
the results of the Gallart et al. (1999b) paper referenced in the text.
The best model representing the data is one
with the SFR(t) in the inset, Z=0.0004, binary fraction 
$f=0.3$, mass ratios of binary stars $q \ge$ 0.7, and flat IMF for the 
secondaries. The IMF is that of Kroupa et al. (1993), but with $x=-3.2$ for 
$m \ge 0.5 M_\odot$.}
\label{leoi}
\end{figure}

\section{How well can we do with CMDs not reaching the oldest MS turnoffs?}

Here I will distinguish two cases depending on the information 
they provide: when the CMD reaches below the HB level, $M_I \simeq 1$, and 
when only 2-3 magnitudes below the tip of the RGB 
are observed, $M_I \simeq -2$ (see Figure~1). In both cases, the CMD 
is populated by stars of all ages, and therefore, it must be possible, in 
principle, to retrieve information about the complete SFH. 
As before, the analysis of the CMD using synthetic CMDs is potentially 
the best method to derive quantitative information. However, 
the detail with which the SFH can be retrieved is necessarily smaller
than in the previous case. 

We are undertaking a project, using CTIO MOSAIC data of the LMC,
to quantify the maximum amount of information on the SFH that one should 
aim to retrieve from CMDs with different magnitude limits. Thanks to its 
proximity, the LMC is ideal for this project because one can obtain 
wide field, good precision old MS turnoff photometry from the ground.
Photometry over large areas is necessary to obtain good statistics of 
stars in short lived evolutionary phases, which are
key in the analysis of shallower CMDs. The aim is to independently retrieve 
the SFH from CMDs of different limiting magnitudes, to learn 
both about the information we loose at each step and about possible 
strategies to help make the most of the available data. Such test is 
important  because we can potentially obtain HB and particularly, RGB 
photometry out to distances like those of the nearest clusters of 
galaxies, and we must know how confidently we can derive SFHs there.
A similar project, to compare SFHs obtained through CMD analysis and 
integrated spectroscopy is underway together with several participants 
in this conference (Alloin, Demarque, Fritze von-Alvensleben, Hardy et al.)

\subsection{CMDs reaching the HB level}

The main age markers here are the (blue and red) HB and the red-clump 
(RC) of core He-burning stars. The presence and appearance of
these features, and in particular that of the HB depends on both age 
and metallicity, although other factors, 
particularly the amount of mass loss during the RGB phase, 
may also play a role (Lee 1993). Very old, low metallicity stars distribute 
along the HB during the core He-burning phase and, generally speaking, 
produce a red HB when the core He burners are not so old, or more 
metal rich, or both. A RC well populated and extended in luminosity (like
that of Leo I in Figure~1) denotes the presence of a large intermediate-age 
population. In addition, the {\it vertical RC} (Zaritsky \& Lin 1997) is 
populated by few hundred Myr old stars that don't undergo the He-flash 
(Beaulieu \& Sackett 1998; Gallart 1998) 

One should aim at retrieving old and intermediate-age star formation 
rates, integrated over intervals of a few Gyr, by fitting the number of 
stars in the HB, the RC and the RGB using synthetic CMDs. In spite of the  
uncertainties in the modeling of these advanced stellar evolution phases, 
the comparison of the models with the
CMDs and luminosity functions of star clusters 
(see Maeder \& Renzini 1983; Renzini \& Fusi Pecci 1988; Chiosi, Bertelli 
\& Bressan 1992) show reasonable agreement, 
(e.g. Zoccali \& Piotto 2000) even though some parameters in the models 
are still uncertain (e.g. the He content, the \cdo or the details
of the mixing processes, see Zoccali et al. 2000). 

HST has provided HB-RC CMDs for a sample of Local Group galaxies outside
the Milky Way system. The M31 dSph satellites have been studied 
by Da Costa et al. (1996, 2000), while the remaining M31 satellites have
been studied by a variety of authors (e.g., M32: Grillmair et al. 1996; 
NGC147: Han et al. 1997; NGC185 and NGC205: Geisler et al. 1998). Similar
data have been analized for some of the dIr 
(Leo A: Tolstoy et al. 1998; 
Pegasus: Gallagher et al. 1998; IC1613: Cole et al. 1999; WLM: Dolphin 2000; 
Sextans A: Dohm-Palmer et al. 1997). Also VLT 
is starting to produce results in this regard (Tolstoy et al. 2000; Held et al. 2000). It is beyond the scope of this paper 
to summarize the results of these works, and I would just like to comment
on very general aspects. In the case of the M31 companions, the main 
general conclusion may be the diversity of their SFHs, much like among 
the Milky Way satellites (Da Costa et al. 2000). For the dIr, it has 
been possible to retrieve in detail the SFH up to 
about 1 Gyr ago using MS and blue-loop stars (Dohm-Palmer et al. 1997), 
and to reveal a prominent intermediate-age star formation.  However, the 
positional overlap of the blue HB with the generally well populated 
$\simeq$ 1 Gyr old MS in the CMD difficults the characterization of the old 
population, and therefore,  additional constraints on the
old stars, like studies of RR Lyrae variables, are important. As 
for the Milky Way satellites, CMDs reaching the oldest MS turnoffs of 
dwarf galaxies through the Local Group would reveal the detail of their 
SFHs and would allow us
to study the early evolution of isolated dwarf galaxies, as well as
dwarfs in a different satellite system like that of M31. And they 
would provide us, again and very likely, with 
fundamental 
surprises. This will become technically feasible for the first time with 
ACS on HST.

Finally, I would like to mention other potentially very useful 
stellar population markers 
in the HB-RC area of the CMD, namely the RGB-bump, the AGB-bump (Gallart 
1998) and the {\it vertical structure} of the RC (Piatti et al. 1999). 
These structures are produced by very short-lived phases of
stellar evolution, and they have only become evident in the CMDs of composite
stellar populations when very densely populated CMDs have been available.
 While the RGB 
and AGB-bumps are produced by stellar populations of basically 
any age, and their structure 
depends on both age and metallicity, the {\it vertical structure} is 
predicted to be produced by relatively metal rich stars (Z $\ge$ 0.004) 
in a very narrow range of masses and ages around 1 Gyr (Girardi 1999). 

\subsection {CMDs reaching just below the tip of the RGB}
 
%In a  CMD of a composite stellar population reaching a magnitude 
%$M_I \simeq -2$, MS and blue-loop stars up to few hundred million years
%old populate the usually called {\it blue-plume}, and red supergiants are
%distributed along a band in the upper red part of the diagram. Older stars 
%are present in a crowded clump in the lower red part of the CMD --the 
%{\it red-tangle}-- and also redwards and above this structure, forming the 
%{\it red-tail} (Gallart, Aparicio \& V\'\i lchez 1996a). The {\it red-tangle}
%is populated by old and intermediate-age RGB and AGB stars in basically the
%whole metallicity range present in the galaxy and, in its bluest region,
%by several hundred Myr old blue-loops. The {\it red-tail} is composed by
%intermediate-age AGB stars of not-so-low metallicities -the higher
%the metallicity, the more extended to the red the {\it red-tail} is.

Gallart et al. (1996a,b) and Aparicio, Gallart \& Bertelli (1997a,b) used 
synthetic CMDs to constrain the SFHs of the dIr galaxies NGC6822,
Pegasus and LGS3 using this kind of data. Only star formation rates
averaged over relatively long periods of time can be obtained in this 
way (see also Aparicio 1998), but it
was also possible to conclude that star formation began in these galaxies
at an early epoch ($\simeq 15-10$ Gyr ago), and that they 
contained a relatively prominent old population. In addition, some 
constraints on the chemical enrichment history were obtained by 
including different parameterizations of Z(t) in the models. 

The position and width of the RGB have been widely used to derive 
the mean metallicity and metallicity dispersion of the stars in dwarf 
galaxies  (Da Costa \& Armandroff 1990; Lee, Freedman \& Madore 1993; 
Saviane et al. 2000). This method, however, assumes 
that the contribution of the dispersion in age to the width of the RGB 
is basically negligible, and it has to be used with care when there is 
a substantial intermediate-age population since, in this case, 
and specially for low metallicity systems, a substantial fraction of the 
RGB width may be contributed by age. Finally, AGB stars are useful tracers 
of intermediate-age star formation (Gallart et al. 1994).

CMDs of this kind are available for basically all Local Group dwarf galaxies.
Besides the ones referenced above, good examples among those not yet 
superseded by published deeper data are the following. Phoenix: 
Mart\'\i nez-Delgado, Gallart \& Aparicio, 1999; NGC3109: Minniti \& 
Zijlstra 1997; DDO210: Lee et al. 2000; Sag DIG: Karachentsev, Aparicio 
\& Makarova 1999; Antlia: Piersimoni et al. 1999; And VI: Armandroff, Jacoby 
\& Davies 1999; Grebel \& Guhatakurta 1999; Gallart et al. 1998).  
CMDs for very distant galaxies, of types not represented in the Local Group 
have been obtained using HST (e.g. Schulte-Ladbeck et al. 1999; 
Harris, Harris \& Poole 1999). 

%\vspace {-0.3 true cm}

\section{Summary and concluding remarks}

Deep CMDs, in particular those reaching the oldest MS turnoffs, are our 
most valuable tools to retrieve {\it in detail} the SFH of nearby galaxies. 
Spectroscopy is a very useful complement, but, on its own, it is not well 
suited to derive SFR($t$), or even Z($t$) directly; instead, the temporal 
information is beautifully laid out in the CMD. Even if possible for a very
limited volume of the Universe, I have argued that this is a worthwhile task,
key to understand galaxy formation and evolution in general, because we have
representatives of most types of galaxies in our neighborhood, and because
using nearby objects, we can learn what can be obtained, and trusted, from  
integrated properties. And most importantly, it is a really feasible task
using the latest generations of telescopes and instruments, particularly 
HST (specially with ACS) and the new generations of 8--10m class telescopes,
equipped with some kind of adaptative optics capability.

But still a lot remains to be done. Even for some of the nearest 
galaxies (the satellites of the Milky Way) for which beautiful CMDs exist, 
analysis using the full power of current modeling techniques are still
lacking. For the most distant galaxies, the optimal data does not yet exist,
but is under reach of current  
instrumentation. Representations of the SFHs of the more distant LG galaxies 
like those presented by Mateo (1998) or Grebel (1998)
provide an useful, {\it qualitative} view of the SFH of these systems as 
inferred from different stellar tracers but should not be taken literally:
{\bf we still don't know, in any level of detail, the evolutionary 
histories of Local Group galaxies outside the Milky Way neighborhood}, and 
reliable {\it population boxes} (Hodge 1989) for these systems 
in such a level of detail will only be available when deeper data 
(reaching at least the HB level, and preferably the old MS turnoffs) will be 
quantitatively analized. But we are getting there. And we will get there 
faster if we can persuade our colleagues on TAC's that this type of study is 
a key path and ingredient in our quest for the global understanding of galaxy 
formation and evolution.

\acknowledgments

I thank A. Aparicio. J. Gallagher and D. Mardones for a critical 
reading of the manuscript,
and the conference organization for financial support.

\end{document}